%% file: main.tex
\pdfoutput=1
\documentclass[sigconf]{acmart}

\usepackage{algorithmic}
\usepackage{graphicx}
\usepackage{textcomp}
\usepackage{xcolor}
\usepackage{booktabs}
\usepackage{multirow}
\usepackage{adjustbox}
\usepackage{threeparttable}
\usepackage{array}
\usepackage{tabularx}
\newcolumntype{C}{>{\centering\arraybackslash}X}
\newcolumntype{R}{>{\raggedleft\arraybackslash}X}
\usepackage{placeins}
\usepackage{microtype}
\usepackage{float}
\usepackage{enumitem}

\AtBeginDocument{%
  }

\setcopyright{none}
\copyrightyear{}
\acmYear{2027}
\acmDOI{}
\acmConference[WSDM '27]{The 20th ACM International Conference on Web
  Search and Data Mining}{February 15--19, 2027}{Hong Kong}
\acmISBN{}
\begin{document}

\title{OptST: Optimizing Semantic Tokenization for LLM-Enhanced Recommendation}

\author{Hanchong Chen}
\affiliation{%
  \institution{Shenzhen Technology University}
  \city{Shenzhen}
  \country{China}}
\affiliation{%
  \institution{The Hong Kong Polytechnic University}
  \city{Hong Kong}
  \country{China}}
\email{hanchongchen.cs@gmail.com}

\author{Jie Liu}
\affiliation{%
  \institution{Shenzhen Technology University}
  \city{Shenzhen}
  \country{China}}
\email{jio1326417@gmail.com}

\author{Fuyuan Lyu}
\affiliation{%
  \institution{McGill University}
  \city{Montreal}
  \country{Canada}}
\email{fuyuan.lyu@mail.mcgill.ca}

\author{Dugang Liu}
\affiliation{%
  \department{College of Computer Science and Software Engineering}
  \institution{Shenzhen University}
  \city{Shenzhen}
  \country{China}}
\email{dugang.ldg@gmail.com}

\author{Lingjie Li}
\affiliation{%
  \institution{Shenzhen Technology University}
  \city{Shenzhen}
  \country{China}}
\email{lilingjie@sztu.edu.cn}

\author{Xiuqiang He}
\affiliation{%
  \department{School of Artificial Intelligence}
  \institution{Shenzhen Technology University}
  \city{Shenzhen}
  \country{China}}
\email{he.xiuqiang@gmail.com}

\author{Xing Tang}
\affiliation{%
  \institution{Shenzhen Technology University}
  \city{Shenzhen}
  \country{China}}
\email{xing.tang@hotmail.com}

\renewcommand{\shortauthors}{Chen et al.}

\input{section/abstract}



\maketitle

\input{section/introduction}
\input{section/related}
\input{section/formulation}
\input{section/method}
\input{section/experiment}
\input{section/conclusion}
\input{section/ethics}
\bibliographystyle{ACM-Reference-Format}
\bibliography{ref}

\end{document}

%% file: section/abstract.tex
\begin{abstract}
Semantic information can enrich the identifiers and structured fields used by recommender systems. Semantic tokenization provides a structured representation of this information by converting LLM-derived user profiles into discrete tokens. In the common two-stage pipeline, the tokenizer is trained before the recommender. Its token assignments are then treated as fixed input features during recommendation training. As a result, the downstream recommender can learn how to use the assigned indices, but it cannot revise the semantic partition that produced them. Making this partition task-aware is difficult because nearest-code assignment is discrete, and unconstrained adaptation may also distort the geometry learned during pretraining. Moreover, a fixed token depth exposes the same semantic resolution to every interaction.
We thus propose \textsc{OptST} (\textbf{Opt}imizing \textbf{S}emantic \textbf{T}okenization), which makes a pretrained tokenizer responsive to downstream recommendation. To revise token assignments without directly updating the pretrained encoder or base codebooks, \textsc{OptST} learns a low-rank deformation whose drift is limited by an anchor regularizer. Recommendation supervision reaches this transformation through a straight-through estimator, while prediction retains hard nearest-code assignments. Beyond adapting the assignments themselves, \textsc{OptST} selects an interaction-conditioned prefix to control how much residual detail is exposed to the recommender. Experiments on four public datasets show that \textsc{OptST} achieves the highest AUC among conventional and LLM-enhanced baselines. Its improvements also persist across different recommendation backbones. These results establish task-aware semantic tokenization as an effective interface between LLM-derived profiles and downstream recommendation models.
\end{abstract}

%% file: section/introduction.tex
\section{Introduction}
\label{sec:intro}
Modern recommender systems mainly rely on structured features, such as user IDs, item IDs, and interaction histories~\cite{deepfm,dcn,autoint,dcnv2}. These features effectively capture collaborative patterns but contain little explicit semantic information. In practical applications, recommendation decisions often depend on semantic knowledge about item content and high-level user preferences that cannot be reliably recovered from interaction identifiers alone~\cite{kar,userip,LangPTune,LettinGo}. This need becomes particularly important in sparse and cold-start scenarios, where collaborative evidence is limited~\cite{idproxy,saviorrec}. Large language models (LLMs) provide a practical way to introduce such knowledge by distilling item content and historical user behavior into natural-language profiles and dense semantic representations~\cite{survey,enhanced,deepseek}. These representations complement conventional collaborative features with explicit descriptions of users and items, making LLM-derived semantics an increasingly important input modality for modern recommender systems.

\begin{figure}[t]
  \centering
  \includegraphics[width=\linewidth]{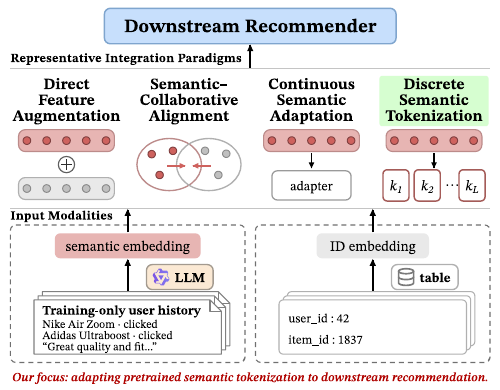}
  \caption{Representative paradigms for integrating LLM-derived semantics
into recommendation models. }
  \Description{Four representative ways of feeding LLM-derived semantics into a
downstream recommender: direct feature augmentation, semantic--collaborative
alignment, continuous semantic adaptation, and discrete semantic tokenization,
the last of which is highlighted as the focus of this work. Below them, two input
modalities: a training-only user history turned into a semantic embedding by an
LLM, and ID features looked up in an embedding table.}
  \label{fig:intro}
\end{figure}

As illustrated in Figure~\ref{fig:intro}, existing methods mainly incorporate LLM-derived semantics into conventional recommenders through three continuous-representation paradigms. The most direct approach treats semantic embeddings as auxiliary features and concatenates them with ID-based representations before feature interaction~\cite{kar,clickprompt}. A second line of work aligns the semantic and collaborative representation spaces, typically via contrastive or matching objectives, making language-derived features more compatible with collaborative behaviors~\cite{ctrl,flip}. A third approach learns lightweight semantic adapters that transform or compress pretrained semantic embeddings before passing them to the downstream recommender, thereby retaining semantic knowledge without involving the LLM during online inference~\cite{ease,llactr}. Despite their different designs, these approaches ultimately represent semantic information as dense, fixed-length continuous vectors. In large-scale applications, caching such vectors for millions of users and items and repeatedly fetching them for long behavior histories or large candidate sets can introduce substantial storage and memory-bandwidth overhead~\cite{better,uist}. Moreover, a flat vector provides neither a compact, shared categorical interface across entities nor an explicit mechanism for accessing semantic information at different levels of resolution. These constraints motivate a storage-efficient semantic interface for downstream recommendation.

Semantic tokenization provides a different interface for delivering semantic information to the downstream recommender, as shown in Figure~\ref{fig:intro}. It maps a dense semantic representation into a short sequence of discrete code indices, commonly using a residual-quantized variational autoencoder (RQ-VAE)~\cite{rqvae}. Semantic tokens were first widely explored as autoregressive identifiers in generative recommendation~\cite{tiger,plum} and have subsequently been introduced as categorical input fields for conventional downstream recommenders via standard embedding lookups~\cite{better,uist}. This interface is independent of how the tokenizer is trained: semantic tokens may be learned through reconstruction alone or further shaped by collaborative alignment and behavior supervision~\cite{das,saviorrec}. Recent systems also combine discrete tokens with continuous semantic features~\cite{RQ-GMM,tokenminds}, showing that these integration interfaces can be complementary. 

Despite these advantages, semantic tokenization is commonly implemented as a two-stage pipeline. A tokenizer is first trained through semantic reconstruction and then used to generate discrete IDs for a downstream recommender~\cite{better,uist,RQ-GMM}. To incorporate recommendation-related information before this boundary, some methods additionally introduce collaborative or behavioral alignment during tokenizer construction~\cite{das,saviorrec}. Such alignment can make the resulting tokens more compatible with collaborative signals, but it remains an upstream proxy for the objective of the downstream recommender. Once the tokens are passed to the recommender, their assignments remain fixed with respect to its training objective. The recommender can learn how to embed and combine the assigned indices, but it cannot revise the quantization partition that produced them. As a result, if two semantic representations are mapped to the same token sequence, their tokenized inputs to the recommender are identical, even when distinguishing them would benefit the downstream objective. The central challenge is therefore to let the downstream recommendation objective revise token assignments while constraining the tokenizer's departure from the semantic structure learned during pretraining.

To address the challenge, we propose \textsc{OptST} (\textbf{Opt}imizing Pretrained \textbf{S}emantic \textbf{T}okenizers), a framework that adapts a pretrained semantic tokenizer with downstream recommendation supervision. Optimizing semantic tokenization requires more than learning new embeddings for fixed token IDs. First, the assignment geometry should capture distinctions that matter to the recommendation while remaining anchored to the semantic structure acquired during pretraining. Second, the residual depth should expose an appropriate level of detail for each user--item interaction rather than applying the same fixed depth universally. These requirements give rise to two main challenges. First, hard nearest-code assignment prevents downstream gradients from reaching the tokenizer, while unconstrained codebook updates may distort the geometry acquired during pretraining. The tokenizer must therefore become task-responsive without drifting into a purely collaborative partition. Second, choosing the token prefix requires discrete continue-or-stop decisions, making it difficult to learn interaction-dependent depth without collapsing to a fixed policy.
\textsc{OptST} addresses these challenges while keeping the pretrained semantic encoder and base RQ codebooks frozen. At each residual layer, it applies a near-identity low-rank transformation shared by all codewords in that layer, allowing the effective assignment geometry to adapt while anchor regularization constrains its deviation from the pretrained codebook. A hard-forward, soft-backward straight-through estimator retains discrete token assignments during prediction while supplying surrogate gradients from the recommendation loss to the low-rank transformations and token embedding tables. \textsc{OptST} further introduces an interaction-conditioned prefix-depth selector that determines how many residual layers to expose based on the current quantization state and interaction context. A straight-through continue-or-stop decision makes the selector trainable, while depth regularization discourages it from defaulting to the maximum depth. The resulting tokens retain the standard categorical-feature interface and can therefore be incorporated into different downstream recommenders without changing their basic architectures. Experiments on four public datasets show that OptST achieves the best performance across all datasets and produces positive gains with five additional recommendation backbones. Ablation studies further verify the effectiveness of codebook adaptation and interaction-conditioned depth selection.
Our main contributions are summarized as follows:
\begin{itemize}[topsep=0pt,noitemsep,nolistsep,leftmargin=*]

\item We study how pretrained hierarchical semantic tokenizers can be adapted for downstream recommendation. We frame this adaptation around two complementary dimensions: the codebook geometry should make token assignments responsive to recommendation signals, while the quantization depth should provide an appropriate semantic resolution for each user--item interaction.

\item We propose \textsc{OptST}, which keeps the pretrained semantic encoder and base codebooks frozen while adapting the effective RQ geometry through shared low-rank transformations, anchor regularization, and straight-through token assignment. An interaction-conditioned depth selector provides complementary control over the resolution exposed to the downstream recommender.

\item We conduct extensive experiments on four public datasets using various recommendation architectures. The results demonstrate improved performance across the evaluated settings and clarify the respective contributions of codebook adaptation and prefix-depth selection.
\end{itemize}

%% file: section/related.tex
\section{Related Work}
\label{sec:related}

\subsection{LLM-Enhanced Recommendation}

Large language models have increasingly been used to complement the collaborative signals learned from user--item interactions with open-world semantic knowledge~\cite{survey,enhanced}. Earlier content-aware recommenders extracted textual features from item descriptions and supplied them directly to prediction models~\cite{baiu}. Recent methods use LLMs to construct richer semantic representations at different granularities, including feature-level knowledge~\cite{kar,llactr}, individual user behaviors~\cite{BAHE}, generated user profiles~\cite{LangPTune,LettinGo}, and multimodal representations for sparse or cold-start entities~\cite{idproxy}. Other studies adapt LLM-derived representations to specific recommendation data through prompting or lightweight semantic modeling~\cite{clickprompt,ease}.
These representations also differ in how they enter the downstream recommender. One approach treats semantic embeddings as auxiliary dense features and combines them with conventional ID representations~\cite{clickprompt,idproxy}. Another aligns semantic and collaborative representation spaces before downstream prediction~\cite{ctrl,flip}. A third introduces lightweight adapters or field-level knowledge injection to transform semantic features into representations compatible with existing recommenders~\cite{ease,kar,llactr}. These mechanisms are not mutually exclusive, but they ultimately expose semantic information through continuous vectors. OptST focuses instead on the discrete semantic interface introduced by semantic tokenization.

\subsection{Semantic Tokenization for Recommendation}

Semantic tokenization maps a dense representation to a short sequence of discrete codes using vector quantization or residual quantization~\cite{vandenoord2017vqvae,rqvae,zeghidour2021soundstream}. In generative recommendation, these codes form the vocabulary through which items are generated. TIGER is an early representative that applies RQ-VAE to item-content embeddings and autoregressively predicts the resulting Semantic IDs~\cite{tiger}. Subsequent methods incorporate collaborative semantics, learnable tokenization, behavior-aware objectives, or contrastive quantization into the construction of Semantic-ID~\cite {lcrec,letter,eager,cost}. PLUM adapts pretrained language models to Semantic-ID generation at an industrial scale, while more recent work studies end-to-end Semantic-ID construction for generative recommendation and advertising~\cite{plum,endtoend}. ETEGRec couples item tokenization and generative recommendation through recommendation-oriented alignment, whereas DIGER allows the generative recommendation objective to influence Semantic-ID learning through differentiable assignments~\cite{etegrec,diger}.

The same discrete interface has also been adopted by discriminative recommenders, where semantic codes enter as categorical fields and are converted into task-specific representations through embedding lookups. Better Generalization with Semantic IDs freezes a content-based RQ-VAE and learns representations of the resulting code sequences in an industrial ranking model~\cite{better}. UIST quantizes user and item semantics and introduces hierarchy-aware interactions between their token sequences~\cite{uist}. UserIP applies vector quantization to LLM-inferred user profiles and couples profile learning with recommendation supervision~\cite{userip}. Other methods introduce behavioral information before or during discretization: QARM aligns multimodal representations before quantization, DAS jointly trains residual quantization and collaborative alignment, and SaviorRec uses fixed RQ assignments to index a trainable behavior-alignment adapter~\cite{qarm,das,saviorrec}. RQ-GMM improves residual discretization through probabilistic distribution modeling, while TokenMinds generates discrete user-interest tokens together with dense user embeddings for production ranking systems~\cite{RQ-GMM,tokenminds}. These studies demonstrate that alignment and tokenization describe different design dimensions: alignment determines how behavioral supervision shapes a representation, whereas tokenization determines how that representation is exposed to the downstream recommender.
Recent studies have also reconsidered how much of a hierarchical token sequence to use. UIST learns soft interactions among different hierarchy levels, whereas TokenMinds adopts a globally selected token prefix~\cite{uist,tokenminds}. In generative recommendation, SA${}^{2}$CRQ, VarLenRec, and CapsID construct variable-length item identifiers using signals such as path entropy, item popularity, or routing confidence~\cite{sa2crq,varlenrec,capsid}. These methods adapt the identifier length at the item level for generative retrieval. OptST instead studies post-pretraining adaptation for a discriminative downstream recommender. It keeps the semantic encoder and base RQ codebooks frozen, uses the downstream recommendation objective to adapt the effective assignment geometry under low-rank and anchor constraints, and selects the token-prefix depth for each user--item interaction.

%% file: section/formulation.tex
\section{Preliminaries}
\label{sec:formulation}

\subsection{Recommendation with Semantic Profiles}

Let $\mathcal{U}$ and $\mathcal{I}$ denote the user and item sets. We consider a dataset $\mathcal{D}=\{(u,i,y_{ui})\}$ of user--item pairs. The label $y_{ui}\in\{0,1\}$ indicates whether $(u,i)$ is a positive or negative recommendation instance under the adopted data construction. Each pair is associated with structured features $\mathbf{x}_{ui}$, including user and item identifiers and other available fields. In addition, an offline LLM-based profiling pipeline converts the available interaction history of user $u$ into a dense semantic profile embedding $\mathbf{p}_u\in\mathbb{R}^{d_p}$. This profiling pipeline remains fixed during downstream training.

Given the structured features and a token-derived semantic representation $\mathbf{s}_{ui}$, a recommender $f_\theta$ predicts a relevance score
\begin{equation}
\hat{y}_{ui}=f_\theta(\mathbf{x}_{ui},\mathbf{s}_{ui})\in(0,1).
\end{equation}
Candidate items can then be ranked according to $\hat{y}_{ui}$. We train the model using binary cross-entropy:
\begin{equation}
\mathcal{L}_{\mathrm{rec}}=-\frac{1}{|\mathcal{D}|}\sum_{(u,i,y_{ui})\in\mathcal{D}}\left[y_{ui}\log\hat{y}_{ui}+(1-y_{ui})\log(1-\hat{y}_{ui})\right].
\label{eq:bce}
\end{equation}

\subsection{Pretrained Residual Semantic Tokenizer}

A residual-quantized variational autoencoder (RQ-VAE)~\cite{vandenoord2017vqvae,rqvae} is pretrained to reconstruct the semantic profile embeddings. Its encoder $g_{\phi_0}$ maps each profile embedding to a latent representation $\mathbf{z}_u=g_{\phi_0}(\mathbf{p}_u)\in\mathbb{R}^{d}$. The tokenizer contains $L$ base codebooks $\mathcal{C}^{0}=\{\mathbf{C}^{0}_{\ell}\}_{\ell=1}^{L}$, where $\mathbf{C}^{0}_{\ell}=\{\mathbf{c}^{0}_{\ell,1},\ldots,\mathbf{c}^{0}_{\ell,K}\}$ contains $K$ codewords at layer $\ell$. After pretraining, the encoder and base codebooks remain frozen. The decoder is not used during downstream recommendation.

Starting from $\mathbf{r}^{0}_{u,0}=\mathbf{z}_u$, residual quantization recursively selects a codeword and updates the remaining residual:
\begin{equation}
k^{0}_{u,\ell}=\arg\min_{k\in\{1,\ldots,K\}}\left\|\mathbf{r}^{0}_{u,\ell-1}-\mathbf{c}^{0}_{\ell,k}\right\|_2^2,\qquad \mathbf{r}^{0}_{u,\ell}=\mathbf{r}^{0}_{u,\ell-1}-\mathbf{c}^{0}_{\ell,k^{0}_{u,\ell}}.
\label{eq:rq}
\end{equation}
The resulting sequence $\mathbf{k}^{0}_u=(k^{0}_{u,1},\ldots,k^{0}_{u,L})$ forms the semantic tokens of user $u$. Successive layers encode the residual left by earlier layers and therefore provide progressively finer information.

The RQ codebooks and the embedding tables used by the downstream recommender have different roles. A codeword $\mathbf{c}^{0}_{\ell,k}\in\mathbb{R}^{d}$ lies in the tokenizer latent space and determines which index is assigned to a profile. A separate trainable table $\mathbf{E}_{\ell}\in\mathbb{R}^{K\times d_s}$ maps the assigned index to the feature $\mathbf{E}_{\ell}[k]$ consumed by the recommender. Updating $\mathbf{E}_{\ell}$ changes how an assigned token is represented downstream, but it does not change which profiles receive that token.

\subsection{Problem Formulation}

Our goal is to adapt a pretrained semantic tokenizer using downstream recommendation supervision while retaining its pretrained semantic structure. The encoder $g_{\phi_0}$ and base codebooks $\mathcal{C}^{0}$ remain frozen. Let $\boldsymbol{\varphi}$ denote the trainable parameters used to construct the effective codebooks $\widetilde{\mathcal{C}}(\boldsymbol{\varphi})$ around $\mathcal{C}^{0}$. Replacing the base codebooks in Equation~\eqref{eq:rq} with the effective codebooks produces the adapted user-level assignment $k_{u,\ell}(\boldsymbol{\varphi})$ at layer $\ell$.

The semantic resolution used by the recommender may depend on the current user--item pair. A gate with parameters $\boldsymbol{\omega}$ therefore selects a prefix of the token sequence. Let $\gamma_{ui,\ell}\in\{0,1\}$ indicate whether layer $\ell$ is included for pair $(u,i)$. We impose $\gamma_{ui,1}=1$ and $\gamma_{ui,\ell+1}\leq\gamma_{ui,\ell}$ so that the selected tokens form a valid prefix. The selected depth $D_{ui}$ and semantic input $\mathbf{s}_{ui}$ are
\begin{equation}
D_{ui}=\sum_{\ell=1}^{L}\gamma_{ui,\ell},\qquad \mathbf{s}_{ui}=\sum_{\ell=1}^{L}\gamma_{ui,\ell}\,\mathbf{E}_{\ell}\!\left[k_{u,\ell}(\boldsymbol{\varphi})\right],
\label{eq:semantic_input}
\end{equation}
where $\mathbf{E}_{\ell}$ is the trainable downstream token embedding table at layer $\ell$.

We optimize the recommender, effective-codebook parameters, token embedding tables, and depth gate through
\begin{equation}
\min_{\theta,\boldsymbol{\varphi},\boldsymbol{\omega},\{\mathbf{E}_{\ell}\}_{\ell=1}^{L}}\ \mathcal{L}_{\mathrm{rec}}+\lambda_a\mathcal{L}_{\mathrm{anchor}}+\lambda_d\mathcal{L}_{\mathrm{depth}},
\label{eq:joint_objective}
\end{equation}
where $\theta$ denotes the recommender parameters, $\mathcal{L}_{\mathrm{anchor}}$ constrains deviation from the frozen base codebooks, and $\mathcal{L}_{\mathrm{depth}}$ penalizes excessive prefix depth. The nonnegative coefficients $\lambda_a$ and $\lambda_d$ control the strengths of the two regularizers. Section~\ref{sec:method} presents the effective-codebook parameterization and the optimization of the discrete assignments and prefix decisions.

%% file: section/method.tex
\section{Methodology}
\label{sec:method}

\begin{figure*}[t]
    \centering
    \includegraphics[width=0.9\textwidth]{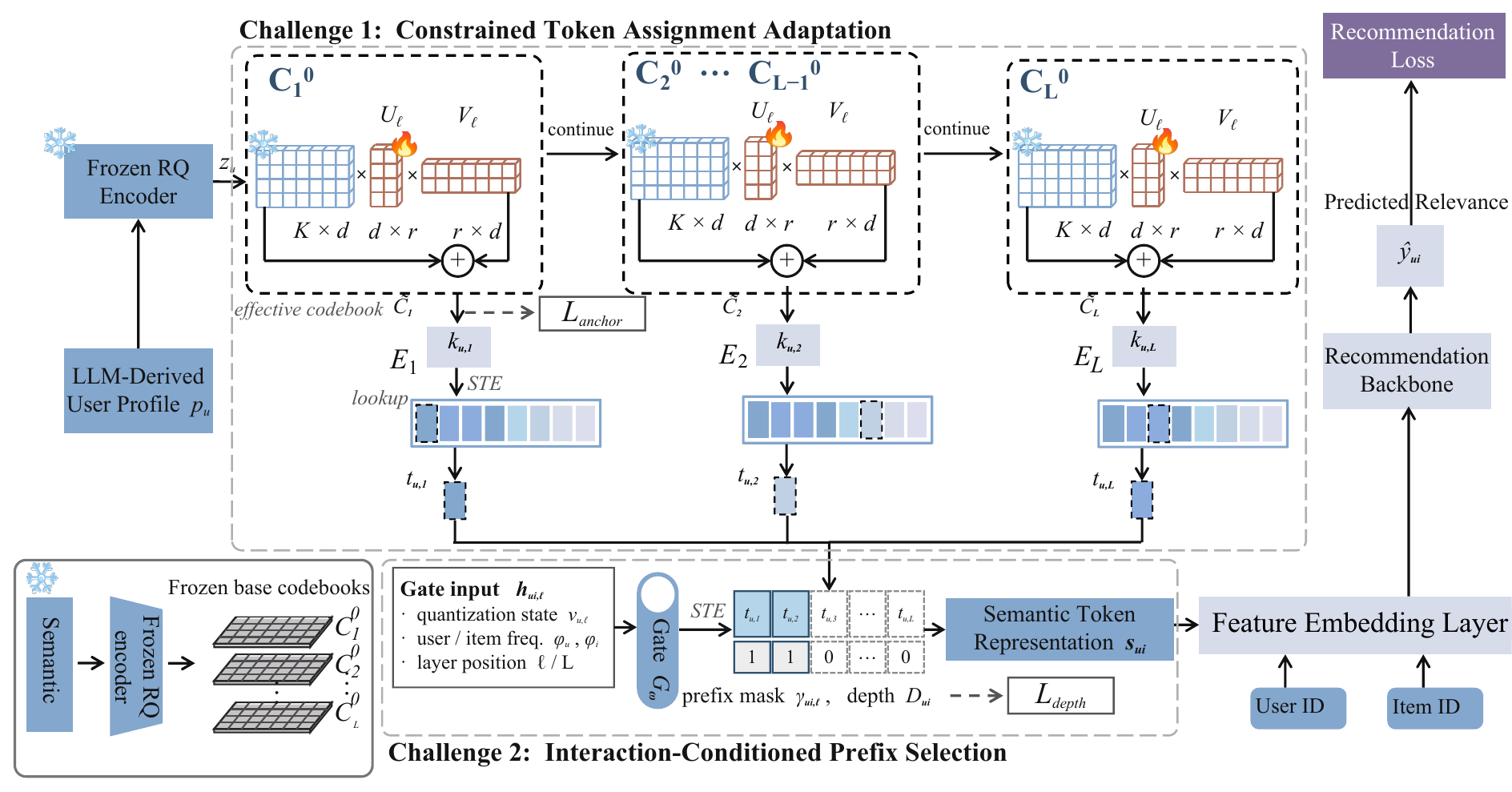}
    \caption{Overview of \textsc{OptST}. Frozen pretrained codebooks anchor token assignment adaptation, while an interaction-conditioned gate selects the token prefix used by the downstream recommender.}
    \Description{Overview of the two components of OptST. Starting from an
    LLM-derived user profile, a frozen semantic encoder and a frozen residual
    quantization encoder produce the user representation. At each residual
    layer, a shared low-rank transformation constructs an effective codebook
    around the frozen base codebook. Hard nearest-code assignment selects a
    token index, while a straight-through estimator propagates recommendation
    gradients and an anchor loss constrains codebook deformation. The gate uses
    the current quantization state, user and item frequencies, and layer
    position to produce a valid prefix mask. The selected token embeddings form
    the semantic representation, which is combined with user and item ID
    embeddings by the downstream recommendation backbone to predict relevance.}
    \label{fig:optst}
\end{figure*}

\subsection{Overview}

Figure~\ref{fig:optst} presents \textsc{OptST}. Starting from a frozen pretrained encoder and frozen base codebooks, \textsc{OptST} adapts the residual token assignments and selects the token prefix used for each user--item pair. The selected indices are mapped through separate trainable lookup tables and combined with the structured features in the downstream recommender.

The first challenge is to make the assignment geometry responsive to downstream recommendation supervision while constraining its deviation from the pretrained geometry. \textsc{OptST} constructs each effective codebook through a near-identity low-rank transformation of its frozen base codebook. It retains hard nearest-code assignment in the forward pass and uses a soft relaxation to propagate recommendation gradients in the backward pass. An anchor regularizer further constrains the resulting geometric deformation.

The second challenge is to select how much residual detail is exposed for each user--item interaction. \textsc{OptST} introduces an interaction-conditioned gate that makes a hard stop-or-continue decision after each quantization layer. A straight-through relaxation enables gate optimization, while a depth regularizer discourages selecting unnecessarily long prefixes. Notice that the downstream recommendation loss provides supervision for both adaptation mechanisms. The token lookup tables and recommender are optimized in the same training process.

\subsection{Constrained Token Assignment Adaptation}
\label{sec:assignment}

The token assignments are determined by the geometry of the codebooks in the RQ latent space. Updating every codeword independently would introduce $Kd$ trainable parameters at each layer and allow the effective geometry to move freely from its pretrained initialization. We instead restrict the downstream adaptation through a layer-wise low-rank transformation of the frozen base codebook.

\paragraph{Low-rank effective codebooks.}
Let $\mathbf{C}_{\ell}^{0}\in\mathbb{R}^{K\times d}$ denote the frozen base codebook at layer $\ell$, where $K$ is the number of codewords and $d$ is their dimension. Inspired by low-rank adaptation~\cite{hu2022lora}, we construct the effective codebook as
\begin{equation}
\widetilde{\mathbf{C}}_{\ell}
=
\mathbf{C}_{\ell}^{0}
+
\mathbf{C}_{\ell}^{0}\mathbf{U}_{\ell}\mathbf{V}_{\ell}
=
\mathbf{C}_{\ell}^{0}
\left(\mathbf{I}_{d}+\mathbf{U}_{\ell}\mathbf{V}_{\ell}\right),
\label{eq:lora}
\end{equation}
where $\mathbf{U}_{\ell}\in\mathbb{R}^{d\times r}$ and $\mathbf{V}_{\ell}\in\mathbb{R}^{r\times d}$ are trainable factors, $r\ll d$ is the adaptation rank, and $\mathbf{I}_{d}\in\mathbb{R}^{d\times d}$ is the identity matrix. The collection $\boldsymbol{\varphi}=\{\mathbf{U}_{\ell},\mathbf{V}_{\ell}\}_{\ell=1}^{L}$ corresponds to the effective-codebook parameters introduced in Section~\ref{sec:formulation}.

We initialize $\mathbf{V}_{\ell}$ to zero, giving $\widetilde{\mathbf{C}}_{\ell}=\mathbf{C}_{\ell}^{0}$ at the beginning of downstream training. The same transformation is applied to all codewords within a layer, so the adaptation introduces only $2dr$ trainable parameters per layer, independently of $K$. This shared parameterization couples the movement of the codewords and biases the update toward a coherent deformation of the layer-wise assignment geometry.

\paragraph{Geometry anchoring.}
The shared low-rank transformation restricts the number of trainable parameters, but it does not directly bound how far the effective codebooks may move during recommendation training. We therefore regularize the effective geometry with respect to the frozen base codebooks. Let $\widetilde{\mathbf{c}}_{\ell,k}$ and $\mathbf{c}_{\ell,k}^{0}$ denote the $k$-th codewords in $\widetilde{\mathbf{C}}_{\ell}$ and $\mathbf{C}_{\ell}^{0}$, respectively. Their pairwise cosine similarities are defined as
\begin{equation}
\widetilde{\rho}_{\ell,km}
=
\cos\!\left(\widetilde{\mathbf{c}}_{\ell,k},\widetilde{\mathbf{c}}_{\ell,m}\right),
\qquad
\rho_{\ell,km}^{0}
=
\cos\!\left(\mathbf{c}_{\ell,k}^{0},\mathbf{c}_{\ell,m}^{0}\right),
\end{equation}
where $\cos(\cdot,\cdot)$ denotes cosine similarity.

For each codeword $k$, let $\mathcal{N}_{\ell,k}^{(J)}$ contain its $J$ nearest neighbors computed once from the frozen codebook at layer $\ell$. We define the anchor loss as
\begin{equation}
\mathcal{L}_{\mathrm{anchor}}
=
\frac{1}{L}\sum_{\ell=1}^{L}
\left[
\frac{\left\|\widetilde{\mathbf{C}}_{\ell}-\mathbf{C}_{\ell}^{0}\right\|_{F}^{2}}{Kd}
+
\frac{1}{KJ}
\sum_{k=1}^{K}
\sum_{m\in\mathcal{N}_{\ell,k}^{(J)}}
\left[
\widetilde{\rho}_{\ell,km}
-
\rho_{\ell,km}^{0}
-
\epsilon_a
\right]_{+}
\right],
\label{eq:anchor}
\end{equation}
where $\|\cdot\|_{F}$ denotes the Frobenius norm, $J$ is the number of pretrained neighbors considered for each codeword, $\epsilon_a\geq0$ is a similarity margin, and $[x]_{+}=\max(0,x)$ is the positive-part operator.

The first term limits the overall displacement of the effective codebooks. The second term penalizes a pretrained neighbor pair only when its cosine similarity increases beyond the margin, thereby discouraging excessive local contraction. This regularizer does not force the pretrained geometry to remain unchanged; it keeps the frozen codebooks as explicit references while allowing recommendation supervision to adjust the assignment boundaries.

\paragraph{Straight-through assignment.}
The effective codebooks can respond to recommendation supervision only if the training signal reflects how the assignment boundaries should change. However, nearest-code selection returns a discrete index and provides no ordinary gradient that compares the selected codeword with alternative assignments. We address this issue with a straight-through estimator that preserves hard tokenization during the forward pass and uses soft assignment only during backward optimization.

At layer $\ell$, let $\mathbf{r}_{u,\ell-1}\in\mathbb{R}^{d}$ denote the residual of user $u$ before quantization. We compute the distance from this residual to every effective codeword as
\begin{equation}
\delta_{u,\ell,k}
=
\left\|
\mathbf{r}_{u,\ell-1}
-
\widetilde{\mathbf{c}}_{\ell,k}
\right\|_{2},
\qquad k\in\{1,\ldots,K\},
\label{eq:assignment_distance}
\end{equation}
where $\widetilde{\mathbf{c}}_{\ell,k}$ is the $k$-th codeword in the effective codebook $\widetilde{\mathbf{C}}_{\ell}$. The hard assignment and residual update are
\begin{equation}
k_{u,\ell}
=
\arg\min_{k\in\{1,\ldots,K\}}
\delta_{u,\ell,k},
\qquad
\mathbf{r}_{u,\ell}
=
\mathbf{r}_{u,\ell-1}
-
\widetilde{\mathbf{c}}_{\ell,k_{u,\ell}}.
\label{eq:hard_assignment}
\end{equation}

To obtain a differentiable surrogate, we normalize the negative distances over all $K$ codewords:
\begin{equation}
q_{u,\ell,k}
=
\frac{
\exp\!\left(-\delta_{u,\ell,k}/\tau\right)
}{
\sum_{j=1}^{K}
\exp\!\left(-\delta_{u,\ell,j}/\tau\right)
},
\label{eq:soft_assignment}
\end{equation}
where $q_{u,\ell,k}$ is the soft assignment weight of codeword $k$ and $\tau>0$ is the assignment temperature. Each layer maintains a trainable lookup table $\mathbf{E}_{\ell}\in\mathbb{R}^{K\times d_s}$, where $d_s$ is the dimension of the token feature used by the recommender. The hard and soft token features are
\begin{equation}
\mathbf{t}^{\mathrm{h}}_{u,\ell}
=
\mathbf{E}_{\ell}[k_{u,\ell}],
\qquad
\overline{\mathbf{t}}_{u,\ell}
=
\sum_{k=1}^{K}
q_{u,\ell,k}\mathbf{E}_{\ell}[k],
\label{eq:hard_soft_features}
\end{equation}
where the superscript $\mathrm{h}$ denotes the hard lookup and $\overline{\mathbf{t}}_{u,\ell}$ denotes its soft relaxation.

The feature passed to the subsequent computation is
\begin{equation}
\widehat{\mathbf{t}}_{u,\ell}
=
\overline{\mathbf{t}}_{u,\ell}
+
\operatorname{sg}\!\left(
\mathbf{t}^{\mathrm{h}}_{u,\ell}
-
\overline{\mathbf{t}}_{u,\ell}
\right),
\label{eq:assignment_ste}
\end{equation}
where $\operatorname{sg}(\cdot)$ is the stop-gradient operator, which acts as the identity in the forward pass and has zero derivative in the backward pass. Consequently, $\widehat{\mathbf{t}}_{u,\ell}$ equals the hard lookup $\mathbf{t}^{\mathrm{h}}_{u,\ell}$ during forward computation, while its backward gradient follows the soft feature $\overline{\mathbf{t}}_{u,\ell}$. Because the soft weights depend on the distances to all effective codewords, the recommendation loss provides a surrogate signal to the assignment geometry and the low-rank factors that construct it.

\subsection{Interaction-Conditioned Prefix Selection}
\label{sec:prefix}

Residual quantization produces an $L$-layer token sequence for each user. Exposing all $L$ layers to every recommendation instance fixes the semantic resolution across user--item pairs. We instead allow the downstream representation to use a pair-specific prefix. The token assignments $k_{u,\ell}$ remain determined by the user profile, while the prefix is conditioned on the current user--item interaction.

\paragraph{Stop-or-continue gate.}
After each layer $\ell\in\{1,\ldots,L-1\}$, a lightweight gating network determines whether the next residual layer should be included. The gate summarizes the current quantization state through
\begin{equation}
\mathbf{v}_{u,\ell}
=
\left[
\left\|\mathbf{r}_{u,\ell}\right\|_2,\;
\log
\frac{\left\|\mathbf{r}_{u,\ell}\right\|_2+\epsilon_{\mathrm{num}}}
{\left\|\mathbf{r}_{u,0}\right\|_2+\epsilon_{\mathrm{num}}},\;
\cos\!\left(
\mathbf{r}_{u,\ell},
\widetilde{\mathbf{c}}_{\ell,k_{u,\ell}}
\right),\;
\mu_{u,\ell}
\right]^{\top},
\label{eq:quantization_state}
\end{equation}
where $\mathbf{v}_{u,\ell}\in\mathbb{R}^{4}$ is the quantization-state vector and $\epsilon_{\mathrm{num}}>0$ is a numerical stabilizer. Its first two components measure the remaining residual and its magnitude relative to the initial latent representation. The third component measures the directional relation between the residual and the selected effective codeword.

The final component $\mu_{u,\ell}$ measures the separation between the two closest codewords:
\begin{equation}
\mu_{u,\ell}
=
\frac{
\delta_{u,\ell}^{(2)}-\delta_{u,\ell}^{(1)}
}{
\delta_{u,\ell}^{(1)}+\epsilon_{\mathrm{num}}
},
\label{eq:assignment_margin}
\end{equation}
where $\delta_{u,\ell}^{(1)}$ and $\delta_{u,\ell}^{(2)}$ are the smallest and second-smallest values among the distances $\{\delta_{u,\ell,k}\}_{k=1}^{K}$ defined in Equation~\eqref{eq:assignment_distance}. A larger $\mu_{u,\ell}$ indicates greater separation between the selected codeword and its nearest alternative.

For either a user or an item, let $x$ denote the entity and let $\mathcal{X}_{\mathrm{tr}}$ denote the corresponding set of training users or items. We define its centered log-frequency feature as
\begin{equation}
\phi_x
=
\log(1+n_x)
-
\mathbb{E}_{x'\sim\mathcal{X}_{\mathrm{tr}}}
\left[\log(1+n_{x'})\right],
\label{eq:frequency_features}
\end{equation}
where $n_x$ is the number of training interactions associated with entity $x$, and the expectation is the empirical mean over $\mathcal{X}_{\mathrm{tr}}$. Substituting $x=u$ and $x=i$ gives the user feature $\phi_u$ and item feature $\phi_i$, respectively.

The complete gate input is
\begin{equation}
\mathbf{h}_{ui,\ell}
=
\left[
\mathbf{v}_{u,\ell}^{\top},\;
\phi_u,\;
\phi_i,\;
\frac{\ell}{L}
\right]^{\top}
\in\mathbb{R}^{7},
\qquad
\mathbf{o}_{ui,\ell}
=
G_{\boldsymbol{\omega}}\!\left(\mathbf{h}_{ui,\ell}\right)
\in\mathbb{R}^{2},
\label{eq:gate_logits}
\end{equation}
where $\ell/L$ identifies the relative layer position, $G_{\boldsymbol{\omega}}$ is the gating network with trainable parameters $\boldsymbol{\omega}$, and $\mathbf{o}_{ui,\ell}$ contains the logits for the stop and continue decisions.

\paragraph{Straight-through gate decisions.}
The stop-or-continue decision is discrete, but replacing it with a purely soft mixture would create a mismatch between training and inference. We therefore retain a hard decision in the forward pass and use a soft relaxation only to approximate its gradient.

Let $o_{ui,\ell}^{\mathrm{stop}}$ and $o_{ui,\ell}^{\mathrm{cont}}$ denote the two entries of the gate logits $\mathbf{o}_{ui,\ell}$. Their soft probabilities are
\begin{equation}
\boldsymbol{\pi}_{ui,\ell}
=
\operatorname{Softmax}
\left(
\frac{\mathbf{o}_{ui,\ell}}{\tau_d}
\right)
=
\left[
\pi_{ui,\ell}^{\mathrm{stop}},
\pi_{ui,\ell}^{\mathrm{cont}}
\right],
\label{eq:gate_probabilities}
\end{equation}
where $\tau_d>0$ is the gate temperature, and $\pi_{ui,\ell}^{\mathrm{stop}}$ and $\pi_{ui,\ell}^{\mathrm{cont}}$ are the probabilities assigned to the two decisions.

The corresponding hard continue indicator is
\begin{equation}
g_{ui,\ell}^{\mathrm{h}}
=
\mathbb{I}
\left[
o_{ui,\ell}^{\mathrm{cont}}
\geq
o_{ui,\ell}^{\mathrm{stop}}
\right],
\label{eq:hard_gate}
\end{equation}
where $\mathbb{I}[\cdot]$ is the indicator function. Thus, $g_{ui,\ell}^{\mathrm{h}}=1$ means that quantization continues to layer $\ell+1$, while $g_{ui,\ell}^{\mathrm{h}}=0$ means that the token sequence stops after layer $\ell$.

We construct the straight-through continue value as
\begin{equation}
\widehat{g}_{ui,\ell}
=
\pi_{ui,\ell}^{\mathrm{cont}}
+
\operatorname{sg}
\left(
g_{ui,\ell}^{\mathrm{h}}
-
\pi_{ui,\ell}^{\mathrm{cont}}
\right).
\label{eq:gate_ste}
\end{equation}
In the forward pass, $\widehat{g}_{ui,\ell}$ equals the binary decision $g_{ui,\ell}^{\mathrm{h}}$. In the backward pass, its gradient follows $\pi_{ui,\ell}^{\mathrm{cont}}$. This allows the recommendation and depth losses to optimize the gate parameters $\boldsymbol{\omega}$ while preserving a discrete token prefix in the forward computation.

\paragraph{Prefix construction and depth regularization.}
The layer-wise continue decisions must form a prefix: once the gate stops, all subsequent token layers are excluded. We construct the binary prefix indicator as
\begin{equation}
\gamma_{ui,1}=1,
\qquad
\gamma_{ui,\ell}
=
\prod_{j=1}^{\ell-1}
g_{ui,j}^{\mathrm{h}},
\quad \ell\in\{2,\ldots,L\},
\label{eq:hard_prefix}
\end{equation}
where $\gamma_{ui,\ell}=1$ indicates that layer $\ell$ is included for pair $(u,i)$. The corresponding hard prefix depth is
\begin{equation}
D_{ui}
=
\sum_{\ell=1}^{L}
\gamma_{ui,\ell},
\label{eq:prefix_depth}
\end{equation}
where $D_{ui}\in\{1,\ldots,L\}$ is the number of token layers exposed to the downstream recommender.

To propagate gradients through the prefix construction, we replace each hard continue indicator in Equation~\eqref{eq:hard_prefix} with its straight-through counterpart:
\begin{equation}
\widehat{\gamma}_{ui,1}=1,
\qquad
\widehat{\gamma}_{ui,\ell}
=
\prod_{j=1}^{\ell-1}
\widehat{g}_{ui,j},
\quad \ell\in\{2,\ldots,L\}.
\label{eq:ste_prefix}
\end{equation}
The forward value of $\widehat{\gamma}_{ui,\ell}$ equals the binary indicator $\gamma_{ui,\ell}$, while its backward gradient follows the relaxed gate probabilities. The semantic representation supplied to the recommender is
\begin{equation}
\mathbf{s}_{ui}
=
\sum_{\ell=1}^{L}
\widehat{\gamma}_{ui,\ell}
\widehat{\mathbf{t}}_{u,\ell},
\label{eq:semantic_representation}
\end{equation}
where $\widehat{\mathbf{t}}_{u,\ell}$ is the straight-through token feature defined in Equation~\eqref{eq:assignment_ste}. In the forward pass, Equation~\eqref{eq:semantic_representation} contains only the hard token lookups within the selected prefix. During training, all $L$ layers are evaluated and their contributions are masked by $\widehat{\gamma}_{ui,\ell}$.

Without an explicit constraint, the gate may default to including every layer. We therefore penalize the normalized prefix depth over a training mini-batch:
\begin{equation}
\mathcal{L}_{\mathrm{depth}}
=
\frac{1}{|\mathcal{B}|}
\sum_{(u,i,y_{ui})\in\mathcal{B}}
\frac{1}{L}
\sum_{\ell=1}^{L}
\widehat{\gamma}_{ui,\ell},
\label{eq:depth_reg}
\end{equation}
where $\mathcal{B}\subseteq\mathcal{D}$ is a mini-batch of recommendation instances and $|\mathcal{B}|$ is its size. In the forward pass, the inner sum equals the hard depth $D_{ui}$. In the backward pass, the straight-through prefix weights provide gradients to the gate. This regularizer discourages the gate from always selecting the full token sequence.

\paragraph{Overall optimization.}
The complete model is trained using the objective in Equation~\eqref{eq:joint_objective}. The recommendation loss is evaluated from the prediction $f_{\theta}(\mathbf{x}_{ui},\mathbf{s}_{ui})$, while the anchor and depth losses are defined in Equations~\eqref{eq:anchor} and~\eqref{eq:depth_reg}. Downstream training updates the low-rank factors $\{\mathbf{U}_{\ell},\mathbf{V}_{\ell}\}_{\ell=1}^{L}$, token lookup tables $\{\mathbf{E}_{\ell}\}_{\ell=1}^{L}$, gate parameters $\boldsymbol{\omega}$, and recommender parameters $\theta$. The pretrained encoder $g_{\phi_0}$ and base codebooks $\{\mathbf{C}_{\ell}^{0}\}_{\ell=1}^{L}$ remain frozen, and the pretrained decoder is not used. The assignment temperature $\tau$ and gate temperature $\tau_d$ control the two gradient relaxations; their settings are reported in Section~\ref{sec:experiment}.

%% file: section/experiment.tex
\section{Experiments}
\label{sec:experiment}

In this section, we conduct experiments to answer the following research questions:
\begin{itemize}[topsep=0pt,noitemsep,nolistsep,leftmargin=*]
\item \textbf{RQ1}: How does \textsc{OptST} compare with conventional and LLM-enhanced recommendation methods?
\item \textbf{RQ2}: Does \textsc{OptST} provide consistent improvements across different downstream recommendation architectures?
\item \textbf{RQ3}: How do constrained token-assignment adaptation and interaction-conditioned prefix selection contribute to the performance of \textsc{OptST}?
\item \textbf{RQ4}: How robust is \textsc{OptST} to the choice of upstream language model?
\item \textbf{RQ5}: How do the learned token assignments and prefix-selection policies behave under different model configurations?
\end{itemize}

\subsection{Experimental Setup}
\label{sec:setup}

\subsubsection{Datasets}

\input{tables/table_stats}
We evaluate \textsc{OptST} on Amazon Movies, Amazon Games, and Amazon Clothing from the Amazon Review dataset\footnote{\url{https://mcauleylab.ucsd.edu/public_datasets/data/amazon/datasets.html}}, together with the Yelp Open Dataset\footnote{\url{https://business.yelp.com/data/resources/open-dataset/}}. Each recorded rating defines an observed user--item interaction $(u,i,r_{ui})$, where $u$, $i$, and $r_{ui}$ denote the user, item, and rating, respectively. Following previous studies~\cite{llactr,ctrl,userip}, we derive a binary preference label as $y_{ui}=\mathbb{I}[r_{ui}>3]$, where $\mathbb{I}[\cdot]$ is the indicator function. Ratings above 3 are treated as positive, while ratings of 3 or less are treated as negative. We retain all recorded interactions without sampling additional negative instances from unobserved user--item pairs.

We adopt the data-splitting protocol used in previous studies~\cite{userip}. Specifically, we use the released split indices, which partition the interactions into training, validation, and test sets with an $8{:}1{:}1$ ratio. All compared methods use the same split indices. Table~\ref{tab:datasets} reports the dataset statistics and training-set positive rates.

\subsubsection{Baselines and Evaluation Metrics}
We compare \textsc{OptST} with thirteen baselines from three groups. The shallow feature-interaction models include FFM~\cite{ffm} and AFM~\cite{afm}. The deep feature-interaction models include AutoInt~\cite{autoint}, DeepFM~\cite{deepfm}, PNN~\cite{pnn}, DCN~\cite{dcn}, DCNv2~\cite{dcnv2}, xDeepFM~\cite{xdeepfm}, and FinalMLP~\cite{finalmlp}. The LLM-enhanced methods include LLaCTR~\cite{llactr}, which introduces field-level semantic knowledge; KAR~\cite{kar}, which augments user and item features with LLM knowledge; CTRL~\cite{ctrl}, which aligns semantic and collaborative representations; and UserIP~\cite{userip}, which quantizes inferred user profiles into collaborative IDs.

Following previous studies~\cite{deepfm,dcn}, we evaluate all methods using AUC and LogLoss. AUC measures the ability to distinguish positive interactions from negative ones, whereas LogLoss evaluates the quality of the predicted probabilities. Higher AUC and lower LogLoss indicate better performance.

\subsubsection{Implementation Details}
We construct semantic profiles after applying the data split. For each user, Qwen3-8B~\cite{qwen3} summarizes the reviews and item attributes associated with training interactions. Validation and test interactions are excluded. The frozen \texttt{all-MiniLM-L6-v2} encoder~\cite{minilm} then maps each summary to a $384$-dimensional profile embedding. Profile generation is performed offline, and the resulting embeddings are cached for downstream experiments. The frequency features used by the depth gate are also computed from the training set. The complete prompts and generation settings are provided with our code.\footnote{\url{https://github.com/chchchchc1023/wsdm27-optst}}
For fair comparison, all LLM-enhanced methods use Qwen3-8B and are evaluated under the same data split, DCN backbone, and hyperparameter search protocol. Beyond these common settings, each baseline retains its original prompting strategy, semantic representation, and integration mechanism. The controlled representation variants are constructed from the same cached profile embeddings.

All models are implemented in PyTorch and optimized with Adam. The common DCN backbone contains three cross layers and two deep layers, each with width $[16,16]$. The ID-embedding dimension is $8$. We search the learning rate and weight decay over $\{10^{-5},5\times10^{-5},10^{-4},5\times10^{-4},10^{-3},5\times10^{-3}\}$. Models are trained with a batch size of $2{,}048$ for at most $100$ epochs, using early stopping on validation AUC with a patience of $5$.
The RQ-VAE encoder is a two-layer MLP with a hidden dimension of $128$ and ReLU activation, and the decoder uses the mirrored architecture. We use $L=8$ quantization layers and select the codebook size $K$ from $\{128,256,512\}$ and the latent dimension $d$ from $\{16,32,64,128\}$. The commitment weight is set to $\beta=0.1$. Each downstream token lookup table has an embedding dimension of $16$, while the depth gate uses a width-$16$ ReLU hidden layer. We tune the low-rank dimension from $\{2,4,8,16\}$, along with the depth and anchor regularization weights, on the validation set. The assignment temperature is fixed at $1.0$, and the gate temperature is annealed from $1.0$ to $0.1$ with a decay rate of $0.95$ per epoch.

\subsection{Overall Performance Comparison (RQ1)}
\label{sec:overall}
\input{tables/table1}

Table~\ref{tab:overall} shows that deep feature-interaction models generally outperform shallow models. This confirms the importance of modeling higher-order relations among recommendation features. LLM-enhanced methods provide further gains, demonstrating that semantic information can complement conventional identifiers and structured fields.

Within the LLM-enhanced group, LLaCTR and UserIP are the strongest baselines. Both establish a tighter connection between semantic information and downstream recommendations. LLaCTR injects field-level semantics into feature interactions, while UserIP represents inferred user profiles as discrete collaborative IDs. \textsc{OptST} achieves the best AUC by going one step further. It allows recommendation supervision to adapt the assignment geometry of a pretrained residual tokenizer. The resulting tokens can preserve distinctions that are useful to the downstream task. The LogLoss results are less uniform, indicating that the primary advantage of \textsc{OptST} lies in discrimination rather than probability calibration.

\subsection{Transferability Study (RQ2)}
\label{sec:transfer}
\input{tables/table_transfer}

To examine whether \textsc{OptST} depends on a particular recommendation architecture, we integrate it with several additional backbones. For each backbone, the downstream adaptation stage is trained separately from the same pretrained tokenizer. Table~\ref{tab:transfer} shows that introducing \textsc{OptST} improves AUC for every backbone across all datasets.

These backbone models feature interactions in different ways. Some rely on attention mechanisms or explicit cross-network connections, while others use factorization structures or multilayer perceptrons. The consistent improvements indicate that task-aware semantic tokenization complements these different interaction mechanisms. The gains also vary across backbone and dataset combinations. This suggests that the useful token partition depends on the downstream model and recommendation data. Therefore, the results demonstrate the broad compatibility of the \textsc{OptST} framework rather than the direct transfer of a single fixed token set.

\subsection{Ablation Studies (RQ3)}
\label{sec:ablation}
\input{tables/table_ablation}
\paragraph{Component and Adaptation Ablations}
Table~\ref{tab:ablation}(a) first compares different semantic representations. Flat quantization does not consistently improve over the continuous profile, showing that discretization alone is insufficient. Frozen RQ-VAE performs better than the continuous representation across all datasets, supporting the use of hierarchical residual tokens. Removing either codebook adaptation or prefix selection reduces AUC. Codebook adaptation generally contributes more, while prefix selection provides a smaller complementary benefit. Anchor regularization improves AUC on Movies and Games, whereas the selected full configurations already use $\lambda_a=0$ on Yelp and Clothing.
Table~\ref{tab:ablation}(b) compares three codebook adaptation strategies while keeping the remaining components unchanged. Free adaptation updates the complete codebook, whereas the additive strategy learns a low-rank residual for individual code entries. The shared strategy instead applies one feature-space transformation to all codewords. It achieves the best AUC with the fewest trainable parameters, supporting the shared transformation as a compact parameterization for downstream adaptation.

\subsection{Sensitivity to the Upstream LLM (RQ4)}
\label{sec:llm}
\input{tables/table_llm_backbone}

Table~\ref{tab:llm} examines the effect of changing the LLM used for profile generation. All remaining components and training settings are kept unchanged. Within the Qwen3 family, larger checkpoints produce gradual AUC improvements across the datasets. The differences are nevertheless limited, showing that \textsc{OptST} does not rely on the largest profile generator to remain effective. Llama-2-7B also performs comparably to the Qwen3 models. This indicates that parameter count alone does not determine the usefulness of the generated profiles. We use Qwen3-8B in the main experiments because it provides the strongest overall point estimates.

\subsection{Model Behavior Analysis (RQ5)}
\label{sec:analysis}

\begin{figure}[!t]
  \centering
  \includegraphics[width=0.245\linewidth]{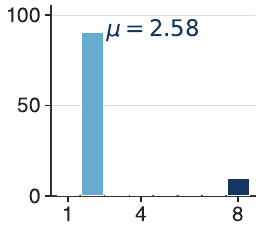}\hfill
  \includegraphics[width=0.245\linewidth]{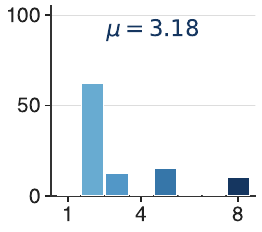}\hfill
  \includegraphics[width=0.245\linewidth]{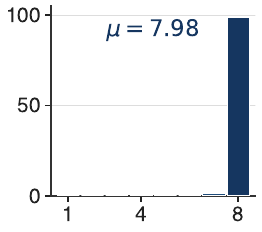}\hfill
  \includegraphics[width=0.245\linewidth]{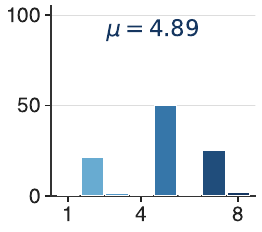}
  \caption{Distribution of prefix depths selected by \textsc{OptST} at test time. Each panel corresponds to one dataset. The horizontal axis denotes prefix depth, the vertical axis denotes the proportion of test interactions, and $\mu$ denotes the mean selected depth.}
  \Description{Four bar charts compare the test-time prefix-depth
  distributions across Movies, Yelp, Games, and Clothing. Movies concentrates
  most selections at depth 2 and has a mean depth of 2.58. Yelp also favors
  depth 2, with smaller groups at depths 3, 5, and 8, giving a mean of 3.18.
  Games selects the full depth of 8 for almost all interactions and has a mean
  of 7.98. Clothing distributes its selections mainly across depths 2, 5, and
  7, with a mean of 4.89. Thus, the learned gate does not impose one universal
  prefix length across datasets.}
  \label{fig:depth}
\end{figure}

\paragraph{Selected Prefix Depth}
Figure~\ref{fig:depth} shows that the learned prefix distributions differ substantially across datasets. Movies and Yelp favor shorter prefixes. Games concentrates near the full token sequence, while Clothing distributes its selections over a wider range of depths. These patterns show that the gate does not impose one universal prefix length. It can expose a short prefix when coarse token information is sufficient and retain deeper residual tokens when the learned recommendation objective favors them.

\begin{figure}[!t]
  \centering
  \includegraphics[width=0.92\linewidth]{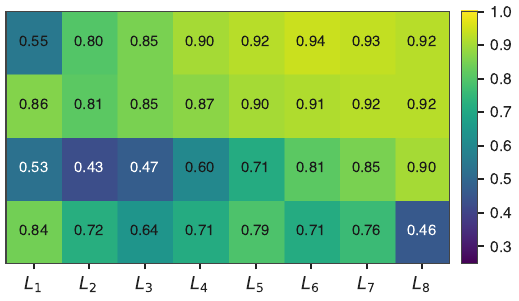}
  \caption{Test-set codebook utilization of \textsc{OptST} with DCN (selection perplexity divided by codebook size). Rows: Movies, Yelp, Games, Clothing; columns: \(L_1\)--\(L_8\).}
  \Description{A four-by-eight heatmap shows normalized selection perplexity
  across residual layers for Movies, Yelp, Games, and Clothing. Higher values
  indicate broader codebook use. Movies increases from 0.55 at the first layer
  to approximately 0.90 or above from the fourth layer onward. Yelp remains
  consistently high, between 0.81 and 0.92. Games starts with lower utilization
  in the first three layers and then rises steadily from 0.60 at layer 4 to
  0.90 at layer 8. Clothing varies across layers and decreases to 0.46 at the
  final layer. Across the four datasets, the assignments remain distributed
  over a substantial portion of the codebook rather than concentrating on only
  a few entries.}
  \label{fig:util}
\end{figure}

\paragraph{Codebook Utilization after Adaptation}
Figure~\ref{fig:util} reports the normalized selection perplexity at each quantization layer. The adapted tokenizer uses a substantial portion of the codebook across datasets and layers. Although the utilization patterns vary, the assignments do not become concentrated on only a few entries. The layer-wise behavior is also nonuniform. Most datasets retain or increase code diversity at deeper residual levels, while Clothing exhibits more concentrated usage in its later layers. This difference is consistent with successive RQ layers serving distinct roles across datasets. Overall, downstream adaptation changes the token assignments without driving the codebook toward a degenerate solution.

%% file: tables/table_stats.tex
\begin{table}[!t]
\centering
\caption{Statistics of the four recommendation datasets. Positive rates are measured on the training partition.}
\label{tab:datasets}
\begin{tabularx}{\linewidth}{@{}l RRRR@{}}
\toprule
\textbf{Dataset} & \textbf{\#Users} & \textbf{\#Items}
  & \textbf{\#Interactions} & \textbf{Pos. Rate} \\
\midrule
Movies   & 7{,}506   & 7{,}360   & 441{,}783      & 71.7\%   \\
Yelp     & 27{,}147  & 20{,}266  & 1{,}293{,}247  & 67.2\%   \\
Games    & 24{,}097  & 10{,}663  & 190{,}043      & 75.1\%   \\
Clothing & 38{,}740  & 22{,}915  & 177{,}537      & 80.6\%   \\
\bottomrule
\end{tabularx}
\end{table}

%% file: tables/table1.tex
\providecommand{\tabbest}[1]{\textbf{#1}}
\providecommand{\tabsnd}[1]{\underline{#1}}
\providecommand{\sigstar}{\textsuperscript{$\ast$}}

\begin{table*}[t]
\centering
\caption{Overall performance on four datasets. The three baseline blocks correspond to shallow feature-interaction, deep feature-interaction, and LLM-enhanced methods, respectively. LL denotes LogLoss. Bold and underlined values indicate the best and second-best results.}
\label{tab:overall}
\small
\renewcommand{\arraystretch}{0.94}
\setlength{\tabcolsep}{3pt}
\begin{tabular*}{0.96\textwidth}{@{\extracolsep{\fill}}lrrrrrrrr@{}}
\toprule
\multirow{2}{*}{\textbf{Method}}
& \multicolumn{2}{c}{\textbf{Movies}}
& \multicolumn{2}{c}{\textbf{Yelp}}
& \multicolumn{2}{c}{\textbf{Games}}
& \multicolumn{2}{c}{\textbf{Clothing}} \\
\cmidrule(lr){2-3}
\cmidrule(lr){4-5}
\cmidrule(lr){6-7}
\cmidrule(lr){8-9}
& AUC$\uparrow$ & LL$\downarrow$
& AUC$\uparrow$ & LL$\downarrow$
& AUC$\uparrow$ & LL$\downarrow$
& AUC$\uparrow$ & LL$\downarrow$ \\
\midrule
FFM
& 0.8144 & 0.4829
& 0.7142 & 0.5968
& 0.6559 & 0.6542
& 0.5521 & 0.7206 \\
AFM
& 0.8079 & 0.4959
& 0.7060 & 0.6206
& 0.6203 & 0.6656
& 0.5419 & 0.7324 \\
\midrule
AutoInt
& 0.8193 & 0.4561
& 0.7384 & 0.5499
& 0.7344 & 0.4977
& 0.6143 & 0.4948 \\
DeepFM
& 0.8187 & 0.4570
& 0.7382 & 0.5502
& 0.7262 & 0.5096
& 0.6143 & 0.5001 \\
PNN
& 0.8182 & 0.4558
& 0.7382 & 0.5505
& 0.7390 & 0.4903
& 0.6141 & 0.5151 \\
DCN
& 0.8195 & 0.4548
& 0.7380 & 0.5507
& 0.7368 & 0.5009
& 0.6160 & 0.5072 \\
DCNv2
& 0.8199 & 0.4533
& 0.7394 & \tabsnd{0.5492}
& 0.7384 & 0.4945
& 0.6195 & 0.5102 \\
xDeepFM
& 0.8208 & 0.4535
& 0.7390 & 0.5498
& 0.7432 & 0.4835
& 0.6175 & 0.5152 \\
FinalMLP
& 0.8159 & 0.4611
& 0.7354 & 0.5534
& 0.7213 & 0.5083
& 0.6048 & 0.5029 \\
\midrule
LLaCTR
& 0.8168 & 0.4610
& \tabsnd{0.7412} & \tabsnd{0.5492}
& 0.7283 & 0.4934
& 0.6279 & 0.5203 \\
KAR
& 0.8056 & 0.4689
& 0.7270 & 0.5590
& 0.7257 & 0.4971
& 0.6194 & \tabsnd{0.4776} \\
CTRL
& 0.8193 & 0.4576
& 0.7375 & 0.5508
& 0.7343 & 0.4894
& 0.6156 & 0.4935 \\
UserIP
& \tabsnd{0.8216} & \tabbest{0.4509}
& 0.7399 & 0.5518
& \tabsnd{0.7488} & \tabsnd{0.4818}
& \tabsnd{0.6324} & \tabbest{0.4734} \\
\midrule
\textsc{OptST}
& \tabbest{0.8237}\sigstar & \tabsnd{0.4510}
& \tabbest{0.7420}\sigstar & \tabbest{0.5482}\sigstar
& \tabbest{0.7532}\sigstar & \tabbest{0.4780}
& \tabbest{0.6380}\sigstar & 0.4879 \\
\bottomrule
\end{tabular*}

\vspace{1pt}
\parbox{0.96\textwidth}{\scriptsize
$\ast$ denotes a statistically significant improvement over the strongest baseline at $p<0.05$. AUC is evaluated using a one-sided DeLong test, while LogLoss is evaluated using a one-sided paired bootstrap test.
}
\end{table*}

%% file: tables/table_transfer.tex
\providecommand{\tabbest}[1]{\textbf{#1}}

\begin{table}[!t]
\centering
\caption{Compatibility of \textsc{OptST} with different recommendation backbones in terms of AUC.}
\label{tab:transfer}
\begin{tabularx}{\linewidth}{@{}l CCCC@{}}
\toprule
\textbf{Method} &
\textbf{Movies} & \textbf{Yelp} & \textbf{Games} & \textbf{Clothing} \\
\midrule
AutoInt
   & 0.8193 & 0.7384 & 0.7344 & 0.6143 \\
\,$+$\textsc{OptST}
   & \tabbest{0.8231} & \tabbest{0.7419} & \tabbest{0.7493} & \tabbest{0.6383} \\
\cmidrule(lr){1-5}
DeepFM
   & 0.8187 & 0.7382 & 0.7262 & 0.6143 \\
\,$+$\textsc{OptST}
   & \tabbest{0.8229} & \tabbest{0.7418} & \tabbest{0.7500} & \tabbest{0.6323} \\
\cmidrule(lr){1-5}
DCNv2
   & 0.8199 & 0.7394 & 0.7384 & 0.6195 \\
\,$+$\textsc{OptST}
   & \tabbest{0.8237} & \tabbest{0.7418} & \tabbest{0.7503} & \tabbest{0.6370} \\
\cmidrule(lr){1-5}
xDeepFM
   & 0.8208 & 0.7390 & 0.7432 & 0.6175 \\
\,$+$\textsc{OptST}
   & \tabbest{0.8239} & \tabbest{0.7422} & \tabbest{0.7508} & \tabbest{0.6380} \\
\cmidrule(lr){1-5}
FinalMLP
   & 0.8159 & 0.7354 & 0.7213 & 0.6048 \\
\,$+$\textsc{OptST}
   & \tabbest{0.8206} & \tabbest{0.7383} & \tabbest{0.7437} & \tabbest{0.6314} \\
\bottomrule
\end{tabularx}
\end{table}

%% file: tables/table_ablation.tex
\providecommand{\tabbest}[1]{\textbf{#1}}
\providecommand{\tabsnd}[1]{\underline{#1}}

\begin{table}[t]
\centering
\caption{Ablations of model components and codebook adaptation strategies in terms of AUC. The w/o Anchor variant sets $\lambda_a=0$. Parentheses in panel (b) report the number of trainable parameters per RQ layer. Bold and underlined values indicate the best and second-best results, respectively.}
\label{tab:ablation}
\footnotesize
\renewcommand{\arraystretch}{0.94}
\setlength{\tabcolsep}{3pt}

\begin{tabularx}{\linewidth}{@{}l CCCC@{}}
\toprule
\multicolumn{5}{c}{\textbf{(a) Component Ablation}} \\
\cmidrule(lr){1-5}
\textbf{Variant}
& \textbf{Movies}
& \textbf{Yelp}
& \textbf{Games}
& \textbf{Clothing} \\
\midrule
Continuous Profile
& 0.8206 & 0.7384 & 0.7449 & 0.6299 \\
\cmidrule(lr){1-5}
Frozen VQ-VAE
& 0.8205 & 0.7406 & 0.7446 & 0.6280 \\
Frozen RQ-VAE
& 0.8219 & 0.7404 & 0.7469 & 0.6366 \\
\cmidrule(lr){1-5}
\textsc{OptST} w/o Adapt.
& 0.8227 & \tabsnd{0.7419}
& 0.7507 & 0.6371 \\
\textsc{OptST} w/o Gate
& \tabsnd{0.8231} & \tabsnd{0.7419}
& \tabsnd{0.7518} & \tabsnd{0.6377} \\
\textsc{OptST} w/o Anchor
& 0.8226 & \tabbest{0.7420}
& 0.7504 & \tabbest{0.6380} \\
\textsc{OptST} (full)
& \tabbest{0.8237} & \tabbest{0.7420}
& \tabbest{0.7532} & \tabbest{0.6380} \\

\midrule
\multicolumn{5}{c}{\textbf{(b) Codebook Adaptation Strategy}} \\
\cmidrule(lr){1-5}
\textbf{Strategy}
& \textbf{Movies}
& \textbf{Yelp}
& \textbf{Games}
& \textbf{Clothing} \\
\midrule
Free (\(Kd\))
& \tabsnd{0.8226} & \tabsnd{0.7419}
& \tabsnd{0.7489} & \tabsnd{0.6371} \\
Additive (\((K+d)r\))
& 0.8225 & 0.7418
& 0.7466 & 0.6369 \\
Shared (\(2dr\))
& \tabbest{0.8237} & \tabbest{0.7420}
& \tabbest{0.7532} & \tabbest{0.6380} \\
\bottomrule
\end{tabularx}
\end{table}

%% file: tables/table_llm_backbone.tex
\providecommand{\tabbest}[1]{\textbf{#1}}
\providecommand{\tabsnd}[1]{\underline{#1}}

\begin{table}[!t]
\centering
\caption{Sensitivity of \textsc{OptST} to the upstream LLM in terms of AUC.}
\label{tab:llm}
\begin{tabularx}{\linewidth}{@{}l CCCC@{}}
\toprule
\textbf{Profile Generator} &
\textbf{Movies} & \textbf{Yelp} & \textbf{Games} & \textbf{Clothing} \\
\midrule
Qwen3-0.6B
   & 0.8230 & 0.7414 & 0.7501 & 0.6365 \\
Qwen3-4B
   & \tabsnd{0.8233} & 0.7416 & 0.7509 & 0.6372 \\
Llama-2-7B
   & \tabsnd{0.8233} & \tabsnd{0.7420} & \tabsnd{0.7511} & \tabsnd{0.6378} \\
Qwen3-8B
   & \tabbest{0.8237} & \tabbest{0.7420} & \tabbest{0.7532} & \tabbest{0.6380} \\
\bottomrule
\end{tabularx}
\end{table}

%% file: section/conclusion.tex
\section{Conclusion}
\label{sec:conclusion}

In this paper, we studied how to optimize a semantic tokenizer for LLM-enhanced recommendation. Rather than treating semantic tokens as fixed upstream features, \textsc{OptST} makes their assignment geometry responsive to downstream recommendation supervision. It applies a shared low-rank transformation around the frozen RQ codebooks and constrains deviation from the pretrained geometry. An interaction-conditioned gate further determines how much residual detail is exposed to the recommender. Experiments on four public datasets show that \textsc{OptST} consistently improves AUC over conventional and LLM-enhanced baselines. Its benefits also persist across different recommendation backbones. The ablation results identify codebook adaptation as the main source of improvement, while prefix selection provides a complementary benefit. These findings support task-aware semantic tokenization as an effective interface between LLM-derived profiles and downstream recommendation models. Future work can extend this framework to temporal profile construction and other semantic modalities.

%% file: section/ethics.tex
\section{Ethical Considerations}
\label{sec:ethics}

All experiments are conducted offline on four public recommendation benchmarks without production deployment or direct interaction with users. The datasets do not contain names, contact information, or explicitly labeled sensitive attributes, though review text and behavioral histories may still reveal personal preferences. The LLM-generated summaries, embeddings, and semantic tokens derived from these data should not be regarded as anonymized, and the upstream language model may reproduce biases present in the source text. Because the benchmarks do not provide protected-group labels, we do not evaluate subgroup fairness in this work.

As with engagement-oriented prediction methods in general, deployment should be mindful of popularity concentration and self-reinforcing feedback loops. Fine-grained user representations could also be repurposed for unintended profiling if access is not properly controlled. Responsible deployment would therefore benefit from subgroup fairness evaluation, profile validation, and appropriate data-governance and user-transparency mechanisms.